\documentclass{aa}
\usepackage{natbib}
\bibpunct{(}{)}{;}{a}{}{,}

\usepackage[dvips]{graphicx}
\usepackage{times}
\usepackage{rotating}
\usepackage{longtable}
\newcommand\ignore[1]{} 

\begin{document}
\title{How many Hipparcos Variability-Induced Movers are genuine binaries?\thanks{Based on observations from the Hipparcos astrometric satellite operated by the European Space Agency (ESA 1997)}}
\titlerunning{How many Hipparcos VIM are genuine binaries?} 
\author{D.~Pourbaix\inst{1,2}\fnmsep\thanks{Research Associate, F.N.R.S., Belgium}
\and 
I.~Platais\inst{1,3,4}
\and
S.~Detournay\inst{1,5}
\and
A.~Jorissen\inst{1}\fnmsep$^{\star\star}$
\and
G.~Knapp\inst{2}
\and
V.V.~Makarov\inst{3,6}}
\institute{
Institut d'Astronomie et d'Astrophysique, Universit\'e Libre de Bruxelles, C.P.~226, Boulevard du Triomphe, B-1050 Bruxelles, Belgium 
\and 
Department of Astrophysical Sciences, Princeton University, Princeton NJ 08543-1001, U.S.A.
\and
Universities Space Research Association, Division of Astronomy and Space Physics, 300 D Street SW, Washington D.C. 20024, U.S.A.
\and
Department of Physics and Astronomy, The Johns Hopkins University, 3400 North Charles Street, Baltimore, MD 21218, U.S.A.
\and
D\'epartement de Physique, Universit\'e de Mons Hainaut, Avenue du Champ de Mars 8, B-7000 Mons, Belgium
\and
US Naval Observatory, Washington D.C., U.S.A.
}

\date{Received date; accepted date} 
\offprints{pourbaix@astro.ulb.ac.be}
\abstract{
Hipparcos observations of some variable stars, and especially of
long-period (e.g. Mira) variables, reveal a motion of the
photocenter correlated with the brightness variation ({\it
  variability-induced mover} -- VIM), suggesting the
presence of a binary companion.  A re-analysis of the Hipparcos
photometric and astrometric data does not confirm the VIM solution for 
62 among the 288 VIM 
objects (21\%) in the Hipparcos catalogue.  Most of these 288 VIMs are 
long-period (e.g. Mira) variables (LPV). 
The effect of a revised chromaticity correction, which accounts for the 
color variations along the light cycle, was then investigated.
It is based on `instantaneous' $V-I$ color indices
derived from Hipparcos and Tycho-2 epoch photometry. 
Among the 188 LPVs flagged as
VIM in the Hipparcos catalogue, 89 (47\%) are not confirmed as VIM
after this improved chromaticity correction is applied. This dramatic
decrease in the number of VIM solutions 
is not surprising, since the chromaticity correction applied
by the Hipparcos reduction consortia was
based on a fixed $V-I$ color. 
Astrophysical considerations lead us to adopt 
a more stringent criterion for accepting a VIM solution 
(first-kind risk of 0.27\% instead of 10\% as in the Hipparcos 
catalogue). With this more severe criterion, only 27 LPV
stars remain VIM, thus rejecting 161 of the 188 (86\%) of the LPVs defined as VIMs in the Hipparcos catalogue.  
}
\maketitle

%
\section{Introduction}
%

The Hipparcos satellite \citep{Hipparcos} obtained high-precision
photometric and astrometric data for some 120\,000 stars, discovering
in the process many new binary systems.  In addition, a small number of 
possible binaries (288) were identified as variability-induced movers
\citep[hereafter VIM,][]{Wielen-1996}.  The physical origin of this
effect lies in a close binary companion of an intrinsic variable star,
causing the photocenter of the system to move as the primary's
luminosity varies.  The Hipparcos analysis allowed for the presence of
such a companion, solving for its position with respect to the
primary in addition to the position, parallax and proper motion of the
system.

Although the VIM model is likely to be appropriate in some cases
\citep{Bertout-1999:a}, the results for some other systems are
questionable.  For instance, rather few VIM have been resolved by speckle 
interferometry even though they have been extensively observed by, for 
instance, the Washington team \citep{Mason-1999:b,Mason-2001:a}.

This situation led us to re-investigate the VIM solutions using
an improved processing of the Hipparcos data.  The entire
available set of astrometric data was used, instead of only FAST 
data, as was the case with the preparation of
the Hipparcos catalogue. This results in 
a decrease by 21\% of the number of VIM solutions, especially the red
ones (Sect.~\ref{Sect:OrigVIM}).

During processing for the Hipparcos catalogue,  corrections were applied to the Hipparcos photometry and
astrometry to correct for the intrinsic chromaticity of the optical system
(i.e. the position of the diffraction spot on the detector depends upon the 
color of the star) and for the aging effects of the  optics and detectors
during the mission.
These effects are especially
important for red variables because of their extreme red colors and
because their variability time scales are comparable to the duration
of the mission. Moreover, red variables have changing colors, and this 
variation must be taken into consideration when applying the
chromaticity correction. In the preparation of the Hipparcos
catalogue, the reduction consortia nevertheless adopted a {\em constant}
color in the chromaticity correction process.
 There is thus room for improvement using chromaticity corrections
 based on {\em epoch} colors, rather than on average colors.
\citet{Platais-2002:a} have devised a scheme to derive epoch colors
based on Tycho-2 and Hipparcos epoch photometry.   
An improved chromaticity correction has then been applied to a sample
of 188 long-period variable stars flagged as VIM in the Hipparcos
catalogue (Sect.~\ref{Sect:NewVIM}), using the epoch $V-I$ index
derived from $Hp - V_{T2}$. Sect.~\ref{Sect:Results} shows that 
a more stringent
criterion (first-kind risk of only 0.27\%) for accepting VIM
solutions seems appropriate based on astrophysical considerations.

%
\section{VIM model and first re-processing}\label{Sect:OrigVIM}
%

Besides the five astrometric parameters ($p_1,\dots,p_5$) used to
model the apparent motion of the center of mass, the VIM model
requires two additional parameters ($D_{\alpha^*},D_{\delta}$) to
provide the direction along which the photocenter moves
\citep{Wielen-1996}.  It is assumed that the companion causing 
the VIM effect has a fixed relative position with respect to the
variable star, i.e. there is no orbital motion on top of the back and
forth  motion of the
photocenter along the segment joining the two
components.  These seven parameters are the unique minimizer of the
least-square problem $\chi^2_V=\Xi^{\mbox{t}}V^{-1}\Xi$
\citep[vol.~1, sect.~2.3.5]{Hipparcos} where
\begin{eqnarray}
\label{Eq:xi}
\Xi&=&\Delta v-\sum_{k=1}^5\frac{\partial v}{\partial p_k}\Delta p_k
\\ \nonumber &&{} -(\frac{\partial v}{\partial
p_1}D_{\alpha^*}+\frac{\partial v}{\partial
p_2}D_{\delta})10^{0.4(Hp_{\rm tot}-Hp_{\rm ref})}
\end{eqnarray}
is a vector with as many components as there are observations, and
$Hp_{\rm tot}$ is the total, instantaneous magnitude of the system,
whereas $Hp_{\rm ref}$ is an arbitrary reference magnitude.  
$\Delta v$ is the abscissa residual along a reference great circle.
The displacements $D_{\alpha^*}$ and $D_{\delta}$ are tied to that
reference magnitude.  Unlike most of the other Hipparcos products, the
VIM solution was built on FAST data only \citep{Lindegren-1997:a}.

Before the effect of the chromaticity correction is investigated in
detail (Sect.~\ref{Sect:NewVIM}), let us first try to reproduce the
original Hipparcos results using the astrometric and
photometric data as they were released with the catalogue.

As far as the Intermediate Astrometric Data (IAD) are concerned,
identifying those which have been used to produce the original VIM
solution is straightforward since
each IAD is flagged with F or N, according to  whether FAST or NDAC used it.
However, not all of them were actually used.  The same rejection criterion as
the one used by the consortia is adopted, namely an abscissa is rejected if its
residual is larger than three times its formal error (as given in field
IA9).

The situation is a bit more obscure for the photometric data given in
the Epoch Photometry Annex.  There might be several observations
corresponding to the very same astrometric data.  Furthermore, there
is a 9-bit flag associated with each observation that tells whether
the magnitude can be trusted, and which consortium used it.  According 
to the Hipparcos explanatory notes \cite[vol.~1, page 220]{Hipparcos},  
if any of bits 3-8 are set, the data are likely to be unreliable. However,
it appears that we cannot strictly follow this
criterion since there is one star (HIP 118188=R Cas) for which all
photometric data are marked as unreliable.  Thus we relaxed this
criterion and accepted a photometric observation  even with bit number 4 set,
i.e. possible interfering object in either field of view.  The median
of the adopted magnitudes is taken as $Hp_{\rm tot}$ for the
corresponding IAD (Eq.~(\ref{Eq:xi})).  
If no magnitude is available for a given IAD, the
reference magnitude ($Hp_{\rm ref}$) is adopted instead.

The statistical significance of the VIM model is assessed through $F_D$:
\begin{equation}
\label{Eq:FD}
F_D^2\!=\!  \left(\!\!
\begin{array}{cc}
D_{\alpha^*} & D_{\delta}
\end{array}
\!\!\right) \left(\!
\begin{array}{cc}
\sigma_{D_{\alpha^*}}^2 &
\rho_D\sigma_{D_{\alpha^*}}\sigma_{D_{\delta}}\\
\rho_D\sigma_{D_{\alpha^*}}\sigma_{D_{\delta}} & \sigma_{D_{\delta}}^2
\end{array}
\!\right)^{-1} \!\!\left(\!\!
\begin{array}{c}
D_{\alpha^*}\\ D_{\delta}
\end{array}
\!\!\right)
\end{equation}
The VIM model is accepted if $F_D>2.15$ which corresponds to a 10\%
chance of a false detection for a Gaussian error distribution. 
This criterion is the same as in the original processing.

\begin{figure*}[htb]
\resizebox{\hsize}{!}{\includegraphics{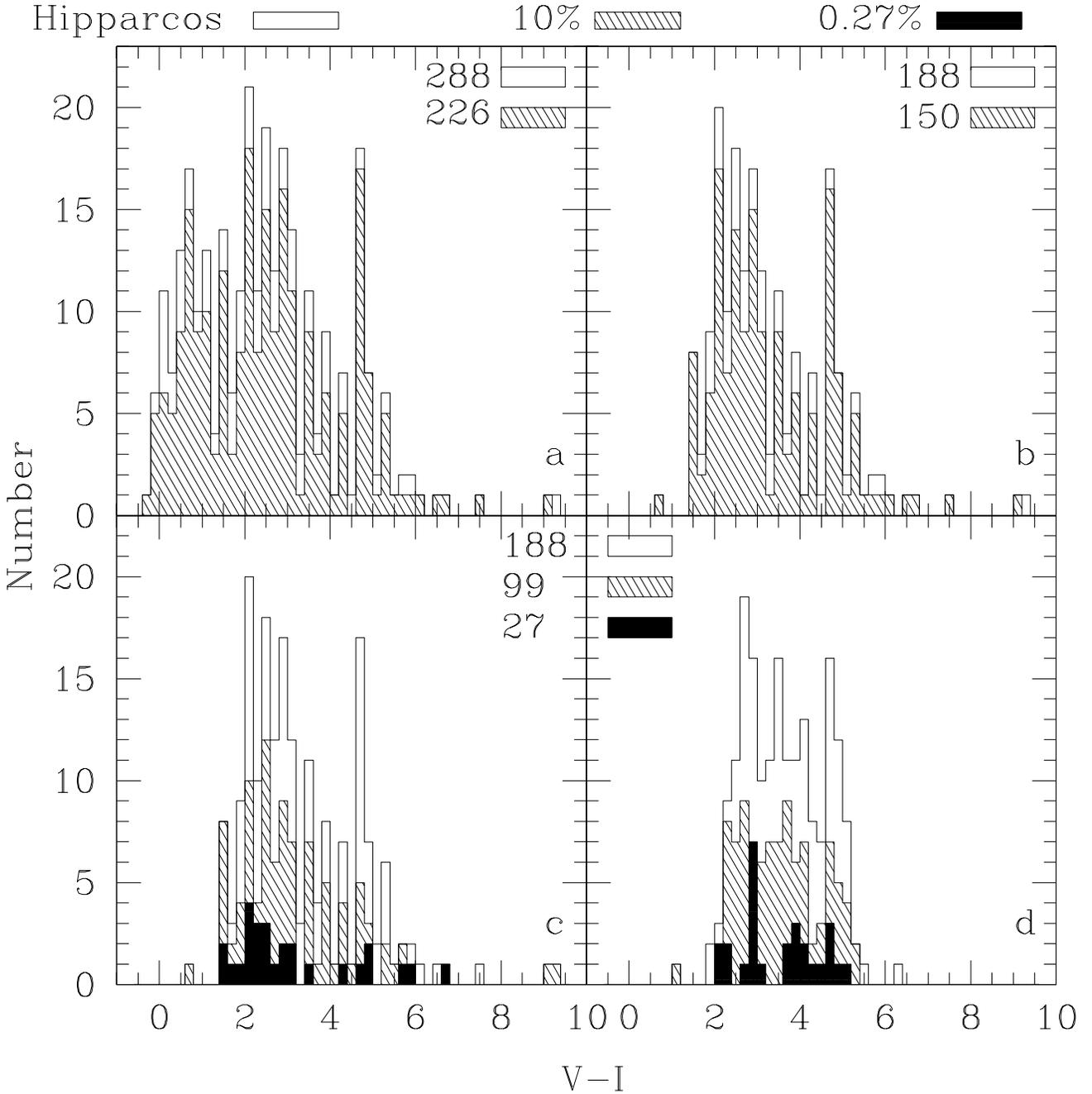}}
\caption[]{\label{Fig:VIM2}{\bf Panel a:} Distribution of the $(V-I)_{H75}$ indices for VIM solutions obtained with a processing similar to that of Hipparcos but applied to the combined FAST \& NDAC data set. The histogram labeled
  `Hipparcos' corresponds to the VIM solutions from the Hipparcos catalogue. 
{\bf Panel b:} Same as (a) for the 188 VIM for which epoch $V-I$ indices are 
available (i.e., for M, S, and C spectral type stars). The VIM solutions are adopted at the 10\% confidence level 
($F_D > 2.15$ in Eq.~(\protect\ref{Eq:FD})). The $V-I$ used in those two panels (both for processing and plotting) is from field H75.
{\bf Panels c \& d:} Same as (b), but after applying a processing 
allowing for time-varying $V-I$ indices and chromaticity
corrections (Sect.~\ref{Sect:NewVIM}). 
The shaded and black histograms correspond to
VIM solutions accepted at the 10\% confidence level ($F_D > 2.15$ in
Eq.~(\protect\ref{Eq:FD})) or 0.27\% confidence level ($F_D > 3.44$), 
respectively.  In panel (c), the histogram is based on H75 whereas the
median of the instantaneous $V-I$ index is used in panel (d).}
\end{figure*}
%

Even though all these criteria are chosen to match as closely as possible those
adopted during the original reduction, the VIM solution based on FAST data is confirmed for only 234 systems, as compared
to 288 in the Hipparcos catalogue (Fig.~\ref{Fig:VIM2}a).  
We checked that the number of accepted VIM solutions depends only
barely upon the criterion used for selecting the photometric data set.
This selection process can thus not be the cause for the decrease in
the number of accepted VIM solutions with our reprocessing. Neither is 
the IAD outlier rejection process, since we checked that the number of 
outlying IAD rejected by the present processing (171/8689 for FAST,
and 189/9050 for NDAC) is not larger than in
the original processing (about 200 for FAST\footnote{This number is
estimated from field H29 ($F1$) giving the fraction of rejected data
points, multiplied by the number of NDAC observations, and summing up
over the whole sample. An inconsistency is sometimes observed in the
catalogue in this respect, since $F1$ may be different from zero, 
although none of the IAD is flagged as rejected!}).  
The only explanation we can think of to account for the larger number
of stars flagged as VIM in the Hipparcos catalogue 
is that the Hipparcos VIM solutions
were derived using slightly different data from that 
published in the catalogue.

If the object is a genuine VIM, the NDAC data should also show the
VIM effect.  Without changing the threshold on $F_D$, one can further
screen the VIM solutions by requiring that the solution based on the
whole data set (FAST \& NDAC) fulfills it.  By doing so, 8
more objects are discarded, resulting in a total decrease of 21\% with
respect to the original number of VIM (Fig.~\ref{Fig:VIM2}a).

%
\section{Revised chromaticity correction}\label{Sect:RCC}
%

Each astrometric measurement is affected by the color of the star,
since it controls the position of the diffraction spot on the
Hipparcos detector \citep[see Sect.~11.3 of
Vol. 3,][]{Hipparcos}. Each star observed by Hip\-par\-cos has
therefore been
assigned a $V-I$ color (or more precisely, $V-I_C$ in the Cousins
system).  This color index is in turn used to compute the chromaticity
correction \citep[Sects.~16.3 and 16.4 of Vol.~3,][]{Hipparcos}.

The two reduction consortia worked independently and concurrently
towards the preparation of the Hipparcos catalogue.  They used
different relations for the chromaticity correction.  NDAC adopted
\citep[Eq.~16.8, vol.~3,][]{Hipparcos}:
\begin{equation}
v^{\rm o}=v^{\rm
c}+[Q_N+\Gamma_{23}+\Gamma_{24}(t-1991.25)]{\lambda_{\rm eff}-550 {\rm
nm}\over 550 {\rm nm}}+\eta
\end{equation}
where $v^{\rm o}$ and $v^{\rm c}$ are the observed and computed
great-circle abscissae, $Q_N$ is the {\em a priori} chromaticity
constant for the
orbit $N$ (kindly provided to us by L. Lindegren), 
$t$ is the epoch of the orbit (expressed in years) and
$\eta$ is the residual noise.  The chromatic parameters are
\begin{eqnarray*}
\Gamma_{23}&=&0.049\pm0.004~{\rm mas},\\
\Gamma_{34}&=&0.010\pm0.005~{\rm mas/yr},
\end{eqnarray*}
$\lambda_{\rm eff}$ is the effective wavelength of a given stellar
spectrum as observed through the $Hp$ bandpass, and can be expressed
as a function of $V-I$.  

FAST accounted for the chromaticity effect
through \cite[Eqs~.16.12--16.18, vol.~3,][]{Hipparcos}:
\begin{eqnarray}
v^{\rm o}&=&v^{\rm c}-0.3221[(V-I)-0.5]\nonumber\\
&&{}+0.1226[(V-I)-0.5]^2\nonumber\\ &&{}-0.0622[(V-I)-0.5](t-1991.25).
\end{eqnarray}
Note that, in the original processing, 
$V-I$ is kept constant in the reduction of any given star
even though the star might be variable, with its color changing
accordingly.  The $V-I$ value used by the consortia for the
astrometric processing is given in field H75 of the Hipparcos
catalogue.

\begin{figure}[htb]
\resizebox{\hsize}{!}{\includegraphics{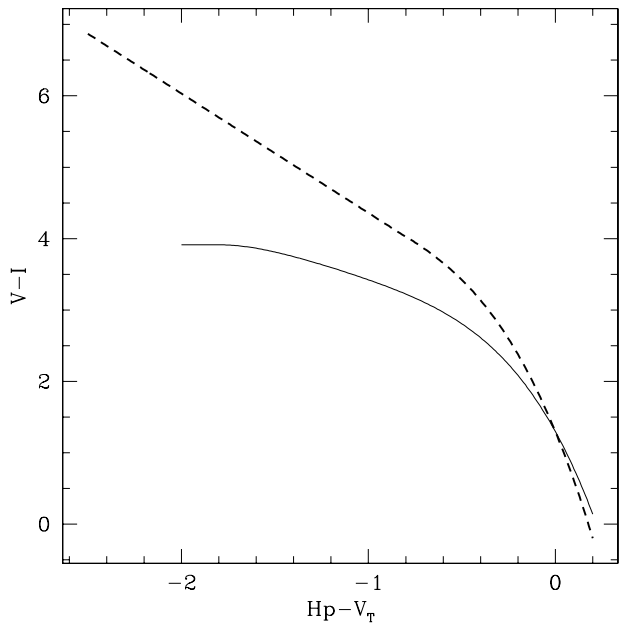}}
\caption[]{\label{Fig:VIcal} The adopted transformation between $Hp-V_T$ and 
$V-I$, specific to carbon stars (solid line) or oxygen stars (dashed
line), according to \citet{Platais-2002:a} }
\end{figure}

Unlike the situation prevailing during the preparation of the
Hipparcos catalogue, now we have access to the Tycho-2 data
\citep{Hog-2000:a}, providing photometry for individual transits, as
well as to recent ground-based $V-I$ observations for selected red
stars.  A detailed analysis of all these photometric data
\citep{Platais-2002:a} yields the transformation relations from
$Hp-V_T$ to $V-I$ given in Fig.~\ref{Fig:VIcal}, which allow us to
find $V-I$ for each epoch $Hp$.

%
\section{New processing}\label{Sect:NewVIM}
%

As mentioned in Sect.~\ref{Sect:RCC}, a constant $V-I$ index was assumed for
each object by the Hipparcos reduction consortia, 
even when this object is a variable star and thus $V-I$ is likely
to change too.  It is therefore possible that many VIM
solutions are just artifacts resulting from the neglected $V-I$
variations.  
In order to check this hypothesis, we reprocessed the VIM stars
and take the $V-I$ variations into account when computing the
chromaticity correction.  The relation between $Hp$ and $V-I$ derived
by \citet{Platais-2002:a} and calibrated
on red stars has been used for that purpose, and, from now on, 
we limit our investigations to those 188
red VIM stars for which the $V-I$ indices can be computed from the data
provided by \citet{Platais-2002:a}.  
To provide an adequate comparison, these 188 red VIM have first been
reprocessed in the Hipparcos manner, as described in
Sect.\ref{Sect:OrigVIM}, with the results displayed in 
Fig.~\ref{Fig:VIM2}b.  Only 150 VIMs are obtained in this Hipparcos-like 
reprocessing.

The results for the reprocessing with variable $V-I$ indices are
presented in Table~\ref{Tab:VIM} and Fig.~\ref{Fig:VIM2}c,d. Only 99 
stars out of 188 (or 53\%) have a VIM solution significant at the 10\% 
level (corresponding to $F_D > 2.15$). This dramatic reduction is a
clear indication that many VIM solutions in the Hipparcos catalogue
are spurious. They result either from the assumed constancy of $V-I$,
as discussed above, or from the inappropriate mean $V-I$ index adopted 
by the Hipparcos reduction consortia (field
H75). Table~\ref{Tab:ReddestVIM} collects 19 stars flagged as VIM in
the Hipparcos catalogue, and with very red $V-I$ indices, namely
$(V-I)_{\rm H75} > 5.0$. Most of these very large values are not confirmed by
the $(V-I, Hp - V_{T2})$ color transformation derived by
\citet{Platais-2002:a}.  Interestingly, that the newly derived $V-I$
indices often agree well with the $(V-I)_{H40}$ indices derived at the 
end of the mission confirms that the $(V-I)_{\rm H75}$ value
adopted by the reduction consortia to derive the chromaticity
correction was incorrect. Among these 19 stars, only 3 (HIP~2215, 
HIP~68815 and HIP~93605) remain VIM objects after our reprocessing.
HIP~93605 (=HD~177017) is indeed a visual binary, with a $V=13.3$
companion located at $1\farcs1$ \citep{Proust-1981:a}  $[$See also Sect.~\ref{Sect:LPVspecific}.$]$.

In Table~\ref{Tab:VIM}, many of the stars flagged as VIM in the Hip\-par\-cos
catalogue, and reprocessed with a single-star solution, keep
almost exactly the same parallax as the Hipparcos one (some examples
are HIP~5559 = RU~Cas, HIP~13502=R~Hor, HIP~25412 = R~Oct, \dots), another 
clear indication that most of the Hipparcos VIM solutions were artifacts. 
In a few cases (HIP~1834, 1901, 65835, 77027, 99082), the Hipparcos parallax 
and the reprocessed parallax are largely different, thus calling for
independent ways to check the new parallaxes. This   is done in the next 
section, where those discrepant cases
are specifically discussed.

Owing to the astrophysical importance of Mira variables,
Table~\ref{Tab:VIM} also provides an estimator of the
quality of the parallax, regardless of the model adopted.
For this purpose, we evaluate not only 
the $\chi^2$ of the adopted model
(denoted $\chi^2_V$, see Eq.~(\ref{Eq:xi})) but also 
the $\chi^2$ (denoted $\chi^2_0)$ for the same model but with
$\varpi = 0$.  The significance of the parallax can then be
evaluated with an F-test:
\begin{equation}
\hat{F}=\frac{N-p}{1}\frac{\chi^2_0-\chi^2_V}{\chi^2_V},
\end{equation}
where $N$ is the number of data points and $p$ the number of
parameters in the model ($p = 5$ for a single-star solution, and $p =
7$ for a VIM solution).  Defining
\begin{equation}
\label{Eq:alpha}
\alpha= Pr\left[\hat{F}<F(1,N-p)|\varpi=0\right],
\end{equation}
a non-zero parallax is considered significant if, say, $\alpha\le5$\%
which roughly corresponds to parallaxes with $\varpi/\sigma_\varpi > 2$.
The value of $\alpha$
is listed in Table \ref{Tab:VIM}.  Eighty objects satisfy the criterion
$\alpha\le5$\%. 

\begin{table*}[htb]
\caption[]{\label{Tab:VIM}188 red VIM from the Hipparcos catalogue, 
for which epoch $(V-I)$ indices are 
  available from the ($V-I, Hp-V_{T2}$) color transformation
  \protect\citep{Platais-2002:a}, reprocessed with an improved
  chromaticity correction. The label `5' 
(standing for 5-parameter, single-star solution) in column `Sol.' means that
  our reprocessing does not confirm the VIM nature of the object, at
  the 0.27\% confidence level ($F_D < 3.44$ in Eq.~(\protect\ref{Eq:FD})).  
The variability type is from \citet{Kholopov-1998:a}. 
$1-\alpha$ is the confidence level that the parallax is different from 
zero (Eq.~(\protect\ref{Eq:alpha})).}
\setlength{\tabcolsep}{0.7mm}
\begin{tabular}{lllcccr|lllcccr}\hline
HIP & GCVS & Var & $\varpi_{\rm HIP}$ (mas) & Sol. & $\varpi$ (mas) &
$\alpha$ & HIP & GCVS & Var & $\varpi_{\rm HIP}$ (mas) & Sol. &
$\varpi$ (mas) & $\alpha$ \\ \hline 
\object{\ignore{HIP }   703} & RU Oct & M & $-2.92\pm2.50$ & 5 & $-1.23\pm1.96$ & 62 & \object{\ignore{HIP } 30449} & V Aur & M & $-1.49\pm3.44$ & 5 & $-0.03\pm1.61$ & 99 \\
\object{\ignore{HIP }   781} & SS Cas & M & $1.40\pm2.33$ & V & $4.98\pm2.02$ & 13 & \object{\ignore{HIP } 31108} & HX Gem & LB: & $12.82\pm3.03$ & V & $11.01\pm2.67$ & 1 \\
\object{\ignore{HIP }  1236} & S Scl & M & $2.13\pm2.65$ & 5 & $2.10\pm1.38$ & 25 & \object{\ignore{HIP } 31484} & TU Aur & L & $2.32\pm1.32$ & 5 & $3.63\pm1.09$ & 4 \\
\object{\ignore{HIP }  1593} & VX And & SRA & $3.56\pm1.28$ & 5 & $2.14\pm0.96$ & 9 & \object{\ignore{HIP } 32115} & RV Pup & M & $1.42\pm1.06$ & 5 & $1.12\pm0.91$ & 33 \\
\object{\ignore{HIP }  1834} & T Cas & M & $0.59\pm1.07$ & 5 & $3.08\pm0.90$ & 2 & \object{\ignore{HIP } 33059} & GY Mon & LB & $3.47\pm1.32$ & 5 & $2.69\pm1.22$ & 11 \\
\object{\ignore{HIP }  1901} & R And & M & $-0.06\pm6.49$ & 5 & $6.96\pm3.63$ & 10 & \object{\ignore{HIP } 34356} & R Gem & M & $-6.22\pm6.50$ & 5 & $-3.71\pm1.56$ & 9 \\
\object{\ignore{HIP }  2180} & AQ And & SR & $0.08\pm1.09$ & 5 & $1.13\pm0.91$ & 42 & \object{\ignore{HIP } 35045} & AA Cam & LB & $1.24\pm1.02$ & 5 & $0.93\pm1.01$ & 48 \\
\object{\ignore{HIP }  2215} & AG Cet & SR & $8.87\pm1.48$ & V & $9.07\pm1.53$ & 0 & \object{\ignore{HIP } 36288} & Y Lyn & L & $4.03\pm1.33$ & V & $4.33\pm1.22$ & 0 \\
\object{\ignore{HIP }  2546} & TU And & M & $-4.26\pm4.43$ & 5 & $0.18\pm2.44$ & 95 & \object{\ignore{HIP } 36314} & VX Aur & M & $0.50\pm3.23$ & 5 & $3.14\pm1.90$ & 28 \\
\object{\ignore{HIP }  2880} & AA Scl & LB & $2.72\pm1.33$ & 5 & $1.15\pm1.18$ & 46 & \object{\ignore{HIP } 36669} & Z Pup & M & $1.33\pm3.73$ & V & $-0.04\pm3.28$ & 99 \\
\object{\ignore{HIP }  4284} & W Cas & M & $-0.17\pm1.61$ & 5 & $0.44\pm1.27$ & 77 & \object{\ignore{HIP } 36675} & S CMi & M & $-4.20\pm4.26$ & 5 & $0.27\pm1.62$ & 91 \\
\object{\ignore{HIP }  4406} & CX Tuc & LB & $1.57\pm1.18$ & 5 & $2.81\pm1.10$ & 2 & \object{\ignore{HIP } 38502} & NQ Pup & L & $3.01\pm1.11$ & 5 & $2.21\pm0.95$ & 6 \\
\object{\ignore{HIP }  5559} & RU Cas & SRB & $3.90\pm1.05$ & 5 & $3.87\pm0.87$ & 1 & \object{\ignore{HIP } 39877} & RT Mon & SRB & $-1.16\pm1.44$ & 5 & $-0.94\pm1.16$ & 55 \\
\object{\ignore{HIP }  6759} & R Scl & SRB & $2.11\pm1.75$ & 5 & $2.06\pm0.80$ & 6 & \object{\ignore{HIP } 40060} & RZ UMa & SRB & $-1.39\pm1.39$ & 5 & $-1.01\pm1.31$ & 53 \\
\object{\ignore{HIP }  6952} & AW Phe & SR & $3.32\pm0.76$ & 5 & $3.45\pm0.67$ & 0 & \object{\ignore{HIP } 40534} & R Cnc & M & $-0.27\pm3.45$ & 5 & $4.39\pm1.39$ & 5 \\
\object{\ignore{HIP }  8025} & V595 Cas & LC & $-0.11\pm1.38$ & 5 & $1.08\pm1.11$ & 50 & \object{\ignore{HIP } 41028} & Z Cnc & SRB & $2.57\pm1.55$ & 5 & $3.10\pm1.26$ & 7 \\
\object{\ignore{HIP }  8034} & V366 And & L & $0.24\pm0.88$ & 5 & $0.76\pm0.78$ & 44 & \object{\ignore{HIP } 41058} & T Lyn & M & $9.97\pm4.33$ & V & $12.21\pm3.97$ & 0 \\
\object{\ignore{HIP }  9767} & Y Eri & M & $3.94\pm1.37$ & 5 & $3.19\pm1.19$ & 2 & \object{\ignore{HIP } 41824} & & UV & $78.05\pm5.69$ & 5 & $85.16\pm6.42$ & 0 \\
\object{\ignore{HIP } 10904} & V605 Cas & LC & $2.36\pm1.36$ & 5 & $1.13\pm1.22$ & 44 & \object{\ignore{HIP } 42975} & R Pyx & M & $0.85\pm1.39$ & 5 & $0.57\pm0.88$ & 63 \\
\object{\ignore{HIP } 11093} & S Per & SRC & $0.62\pm1.88$ & V & $0.56\pm1.76$ & 87 & \object{\ignore{HIP } 43063} & EY Hya & SRA & $1.00\pm1.84$ & 5 & $2.85\pm1.44$ & 27 \\
\object{\ignore{HIP } 11284} & V441 Per & LC & $0.93\pm1.37$ & 5 & $-0.32\pm1.21$ & 82 & \object{\ignore{HIP } 43575} & BO Cnc & LB: & $3.68\pm1.18$ & V & $5.13\pm1.03$ & 0 \\
\object{\ignore{HIP } 11293} & TZ Hor & L & $3.54\pm0.60$ & 5 & $3.76\pm0.54$ & 0 & \object{\ignore{HIP } 43653} & S Hya & M & $-1.26\pm2.71$ & 5 & $-3.04\pm2.48$ & 42 \\
\object{\ignore{HIP } 12193} & R Tri & M & $2.51\pm1.69$ & 5 & $0.33\pm1.23$ & 82 & \object{\ignore{HIP } 44995} & W Cnc & M & $-11.89\pm4.71$ & 5 & $-1.61\pm2.40$ & 68 \\
\object{\ignore{HIP } 12781} & VX Ari & UV & $132.42\pm2.48$ & 5 & $132.75\pm2.09$ & 0 & \object{\ignore{HIP } 45824} & IN Hya & SRB & $2.52\pm1.17$ & 5 & $4.03\pm1.01$ & 0 \\
\object{\ignore{HIP } 13502} & R Hor & M & $3.25\pm1.08$ & 5 & $3.24\pm0.92$ & 2 & \object{\ignore{HIP } 45915} & CG UMa & LB & $6.84\pm1.00$ & V & $7.31\pm0.89$ & 0 \\
\object{\ignore{HIP } 14042} & T Hor & M & $-0.15\pm1.40$ & 5 & $0.69\pm1.22$ & 63 & \object{\ignore{HIP } 46806} & R Car & M & $7.84\pm0.83$ & V & $7.41\pm0.81$ & 0 \\
\object{\ignore{HIP } 14229} & V Hor & SRB & $2.97\pm0.64$ & 5 & $2.44\pm0.58$ & 0 & \object{\ignore{HIP } 47066} & X Hya & M & $0.06\pm2.53$ & 5 & $3.86\pm1.97$ & 17 \\
\object{\ignore{HIP } 15145} & AA Per & SRA & $2.36\pm1.83$ & 5 & $0.85\pm1.64$ & 74 & \object{\ignore{HIP } 47630} & R Sex & L & $2.71\pm1.68$ & 5 & $3.63\pm1.39$ & 4 \\
\object{\ignore{HIP } 15474} & $\tau^4$ Eri & L & $12.63\pm0.89$ & 5 & $11.90\pm0.74$ & 0 & \object{\ignore{HIP } 47886} & R LMi & M & $1.87\pm2.54$ & 5 & $3.36\pm2.18$ & 30 \\
\object{\ignore{HIP } 15926} & VX Eri & SR: & $5.18\pm1.95$ & 5 & $2.61\pm1.20$ & 10 & \object{\ignore{HIP } 48036} & R Leo & M & $9.87\pm2.07$ & 5 & $12.17\pm1.43$ & 0 \\
\object{\ignore{HIP } 16647} & RT Eri & M & $2.15\pm4.13$ & 5 & $3.61\pm1.58$ & 9 & \object{\ignore{HIP } 48662} & X Vel & SR & $1.90\pm0.79$ & 5 & $2.00\pm0.66$ & 3 \\
\object{\ignore{HIP } 18165} & BL Tau & SR: & $3.17\pm2.27$ & 5 & $2.12\pm1.47$ & 22 & \object{\ignore{HIP } 52988} & WW Vel & M & $-1.65\pm3.21$ & 5 & $-6.40\pm3.41$ & 15 \\
\object{\ignore{HIP } 18931} & DP Eri & SRB & $4.93\pm0.91$ & 5 & $3.43\pm0.82$ & 0 & \object{\ignore{HIP } 53085} & V Hya & L & $0.16\pm1.29$ & 5 & $1.18\pm1.16$ & 49 \\
\object{\ignore{HIP } 19931} & SY Per & SRA & $2.30\pm2.44$ & 5 & $5.42\pm2.05$ & 5 & \object{\ignore{HIP } 53809} & R Crt & SRB & $0.45\pm1.78$ & 5 & $4.16\pm1.29$ & 10 \\
\object{\ignore{HIP } 21479} & R Dor & SRB & $16.02\pm0.69$ & 5 & $17.04\pm0.56$ & 0 & \object{\ignore{HIP } 54951} & FN Leo & L & $0.49\pm1.30$ & V & $2.65\pm1.11$ & 2 \\
\object{\ignore{HIP } 21766} & R Cae & M & $-1.83\pm1.46$ & 5 & $-0.33\pm1.11$ & 85 & \object{\ignore{HIP } 57009} & DN Hya & M & $0.08\pm1.80$ & 5 & $0.51\pm1.37$ & 77 \\
\object{\ignore{HIP } 22127} & X Cam & M & $1.01\pm1.22$ & 5 & $1.16\pm1.04$ & 41 & \object{\ignore{HIP } 57642} & X Cen & M & $0.42\pm2.23$ & 5 & $-0.60\pm1.71$ & 80 \\
\object{\ignore{HIP } 22256} & SU Dor & M & $1.74\pm1.44$ & 5 & $0.90\pm1.26$ & 53 & \object{\ignore{HIP } 57917} & S Crt & SRB & $2.04\pm1.31$ & 5 & $1.38\pm1.13$ & 35 \\
\object{\ignore{HIP } 22552} & ST Cam & SRB & $0.36\pm0.84$ & 5 & $-0.04\pm0.70$ & 97 & \object{\ignore{HIP } 59943} & AO Cru & LC & $0.22\pm0.84$ & 5 & $-0.89\pm0.76$ & 30 \\
\object{\ignore{HIP } 22667} & $o^1$ Ori & I & $6.02\pm0.94$ & 5 & $5.25\pm0.81$ & 0 & \object{\ignore{HIP } 60180} & RY UMa & SRB & $2.26\pm0.79$ & 5 & $3.21\pm0.72$ & 0 \\
\object{\ignore{HIP } 22670} & V346 Aur & SRA: & $-2.95\pm1.77$ & 5 & $0.98\pm1.35$ & 65 & \object{\ignore{HIP } 61667} & R Vir & M & $1.40\pm1.41$ & 5 & $2.59\pm0.99$ & 3 \\
\object{\ignore{HIP } 23200} & V1005 Ori & BY+UV & $37.50\pm2.56$ & 5 & $39.13\pm2.35$ & 0 & \object{\ignore{HIP } 62071} & UW Cen & RCB & $1.27\pm2.81$ & 5 & $1.65\pm2.74$ & 66 \\
\object{\ignore{HIP } 23520} & EL Aur & LB & $4.81\pm1.88$ & V & $4.74\pm1.59$ & 1 & \object{\ignore{HIP } 62401} & RU Vir & M & $-2.25\pm2.86$ & 5 & $-4.37\pm2.40$ & 17 \\
\object{\ignore{HIP } 23636} & T Lep & M & $-1.77\pm2.73$ & 5 & $0.89\pm1.40$ & 65 & \object{\ignore{HIP } 62712} & U Vir & M & $1.72\pm1.85$ & 5 & $2.55\pm1.61$ & 31 \\
\object{\ignore{HIP } 23680} & W Ori & SRB & $4.66\pm1.44$ & 5 & $3.45\pm1.04$ & 1 & \object{\ignore{HIP } 63152} & RY Dra & L & $2.05\pm0.65$ & 5 & $2.17\pm0.61$ & 1 \\
\object{\ignore{HIP } 25412} & R Oct & M & $3.28\pm1.23$ & 5 & $3.17\pm0.95$ & 0 & \object{\ignore{HIP } 63950} & FS Com & SRB & $5.70\pm1.19$ & 5 & $3.69\pm0.98$ & 1 \\
\object{\ignore{HIP } 27181} & Y Tau & SRB & $3.72\pm2.29$ & 5 & $1.68\pm0.98$ & 24 & \object{\ignore{HIP } 64768} & FH Vir & SR & $2.18\pm1.04$ & 5 & $2.27\pm0.85$ & 5 \\
\object{\ignore{HIP } 27286} & S Col & M & $2.60\pm2.05$ & 5 & $0.94\pm1.82$ & 70 & \object{\ignore{HIP } 64778} & UY Cen & SR & $1.66\pm1.04$ & 5 & $1.66\pm0.84$ & 16 \\
\object{\ignore{HIP } 27398} & FU Aur & L & $1.86\pm1.69$ & 5 & $-0.55\pm1.31$ & 72 & \object{\ignore{HIP } 65242} & UX Cen & I & $2.51\pm0.96$ & 5 & $2.56\pm0.86$ & 2 \\
\object{\ignore{HIP } 28166} & BO Ori & SR & $4.09\pm1.33$ & 5 & $4.18\pm1.15$ & 3 & \object{\ignore{HIP } 65835} & R Hya & M & $1.62\pm2.43$ & 5 & $8.44\pm1.00$ & 0 \\
\hline
\end{tabular}
\end{table*}

\begin{table*}[htb]
\addtocounter{table}{-1}
\caption[]{(cont.)}
\setlength{\tabcolsep}{0.7mm}
\begin{tabular}{lllcccr|lllcccr}\hline
HIP & GCVS & Var & $\varpi_{\rm HIP}$ (mas) & Sol. & $\varpi$ (mas) &
$\alpha$ & HIP & GCVS & Var & $\varpi_{\rm HIP}$ (mas) & Sol. &
$\varpi$ (mas) & $\alpha$ \\ \hline
\object{\ignore{HIP } 66100} & S Vir & M & $-3.04\pm4.29$ & 5 & $1.47\pm1.44$ & 42 & \object{\ignore{HIP } 94162} & SZ Dra & L & $3.13\pm0.76$ & 5 & $3.02\pm0.69$ & 0 \\
\object{\ignore{HIP } 66466} & RV Cen & M & $1.00\pm1.13$ & 5 & $0.12\pm0.84$ & 92 & \object{\ignore{HIP } 94224} &V3954 Sgr  & LB: & $3.91\pm1.24$ & 5 & $4.10\pm1.04$ & 1 \\
\object{\ignore{HIP } 67410} & R CVn & M & $-0.49\pm1.47$ & V & $-2.33\pm1.27$ & 9 & \object{\ignore{HIP } 94706} & T Sgr & M & $-31.67\pm9.28$ & V & $-16.14\pm8.52$ & 0 \\
\object{\ignore{HIP } 67419} & W Hya & SRA & $8.73\pm1.09$ & 5 & $12.85\pm0.99$ & 0 & \object{\ignore{HIP } 94738} & R Sgr & M & $1.19\pm1.90$ & 5 & $2.59\pm1.45$ & 20 \\
\object{\ignore{HIP } 67626} & RX Cen & M & $-17.92\pm7.60$ & 5 & $-8.59\pm5.51$ & 31 & \object{\ignore{HIP } 95173} & T Sge & L & $2.22\pm1.25$ & 5 & $2.81\pm1.15$ & 8 \\
\object{\ignore{HIP } 68815} & $\theta$ Aps & SRB & $9.93\pm0.64$ & V & $10.36\pm0.63$ & 0 & \object{\ignore{HIP } 95676} & SW Tel & M & $-5.39\pm6.58$ & V & $-5.50\pm5.74$ & 35 \\
\object{\ignore{HIP } 68837} & U Cir & SR & $-1.92\pm1.51$ & 5 & $-2.31\pm1.35$ & 17 & \object{\ignore{HIP } 96836} & TT Cyg & SRB & $1.96\pm0.80$ & 5 & $1.36\pm0.70$ & 14 \\
\object{\ignore{HIP } 70885} & V Boo & SRA & $4.06\pm0.93$ & 5 & $4.86\pm0.83$ & 0 & \object{\ignore{HIP } 97586} & GY Aql & SR & $9.66\pm5.53$ & 5 & $9.60\pm4.26$ & 14 \\
\object{\ignore{HIP } 70969} & Y Cen & I & $3.15\pm1.29$ & 5 & $5.57\pm1.26$ & 0 & \object{\ignore{HIP } 97629} & $\chi$ Cyg & M & $9.43\pm1.36$ & 5 & $6.71\pm1.00$ & 0 \\
\object{\ignore{HIP } 71802} & RW Boo & L & $3.09\pm1.10$ & 5 & $2.48\pm1.00$ & 5 & \object{\ignore{HIP } 97644} & T Pav & M & $-0.54\pm1.43$ & 5 & $1.02\pm1.20$ & 52 \\
\object{\ignore{HIP } 75143} & S CrB & M & $1.90\pm1.36$ & 5 & $2.40\pm1.17$ & 15 & \object{\ignore{HIP } 98031} & S Pav & SRA & $2.91\pm1.77$ & 5 & $3.94\pm1.09$ & 1 \\
\object{\ignore{HIP } 75393} & RS Lib & M & $1.74\pm2.06$ & 5 & $3.99\pm1.78$ & 8 & \object{\ignore{HIP } 98190} & AX Cyg & L & $2.79\pm0.86$ & 5 & $2.02\pm0.70$ & 3 \\
\object{\ignore{HIP } 75727} & GO Lup & SRB & $4.30\pm1.35$ & V & $4.46\pm1.31$ & 0 & \object{\ignore{HIP } 99082} & V1943 Sgr & LB & $1.82\pm1.82$ & 5 & $5.02\pm0.97$ & 0 \\
\object{\ignore{HIP } 76377} & R Nor & M & $5.14\pm1.87$ & V & $2.96\pm1.71$ & 20 & \object{\ignore{HIP } 99653} & RS Cyg & SRA & $1.81\pm0.84$ & V & $2.35\pm0.78$ & 1 \\
\object{\ignore{HIP } 76460} & SY CrB & LB & $4.73\pm1.28$ & 5 & $4.21\pm1.29$ & 1 & \object{\ignore{HIP } 99990} & RT Cap & SRB & $1.78\pm1.48$ & 5 & $2.47\pm0.97$ & 7 \\
\object{\ignore{HIP } 77023} & FQ Lup & L: & $2.83\pm1.50$ & 5 & $3.48\pm1.11$ & 3 & \object{\ignore{HIP }100048} & CN Cyg & M & $-1.63\pm1.34$ & 5 & $-0.70\pm1.19$ & 61 \\
\object{\ignore{HIP } 77027} & BG Ser & M & $-3.78\pm2.74$ & 5 & $2.67\pm2.05$ & 32 & \object{\ignore{HIP }100404} & BC Cyg & L & $2.84\pm0.87$ & V & $3.64\pm0.76$ & 0 \\
\object{\ignore{HIP } 77284} & Y CrB & L & $4.00\pm1.47$ & 5 & $4.77\pm1.38$ & 0 & \object{\ignore{HIP }100582} & V744 Cyg & LB & $2.62\pm2.26$ & 5 & $2.78\pm1.98$ & 32 \\
\object{\ignore{HIP } 78307} & AH Ser & M & $-16.33\pm7.71$ & 5 & $-6.84\pm5.62$ & 34 & \object{\ignore{HIP }100605} & UU Dra & SRB & $2.26\pm0.87$ & 5 & $3.47\pm0.77$ & 0 \\
\object{\ignore{HIP } 78976} & U Ser & M & $2.46\pm3.29$ & 5 & $1.82\pm2.80$ & 59 & \object{\ignore{HIP }101888} & RU Vul & SRA & $0.90\pm1.30$ & 5 & $1.27\pm1.12$ & 29 \\
\object{\ignore{HIP } 79233} & RU Her & M & $1.89\pm1.82$ & V & $0.81\pm1.62$ & 65 & \object{\ignore{HIP }102082} & V Cyg & M & $3.69\pm1.77$ & 5 & $2.27\pm1.26$ & 11 \\
\object{\ignore{HIP } 80259} & RY CrB & SRB & $0.92\pm1.30$ & V & $1.42\pm1.16$ & 31 & \object{\ignore{HIP }102246} & S Del & M & $-0.50\pm1.74$ & 5 & $-0.20\pm1.82$ & 93 \\
\object{\ignore{HIP } 80802} & R UMi & SRB & $1.79\pm0.93$ & 5 & $3.85\pm0.75$ & 0 & \object{\ignore{HIP }102829} & T Aqr & M & $-1.02\pm1.76$ & 5 & $-0.12\pm1.60$ & 95 \\
\object{\ignore{HIP } 81309} & X Ara & M & $-2.04\pm6.69$ & 5 & $-2.30\pm5.46$ & 76 & \object{\ignore{HIP }104451} & T Cep & M & $4.76\pm0.75$ & 5 & $5.96\pm0.59$ & 0 \\
\object{\ignore{HIP } 81506} & AS Her & M & $5.67\pm1.88$ & 5 & $4.93\pm1.56$ & 5 & \object{\ignore{HIP }106583} & S Cep & M & $2.41\pm0.61$ & 5 & $2.42\pm0.51$ & 0 \\
\object{\ignore{HIP } 81747} & AX Sco & SRB & $4.52\pm1.57$ & 5 & $3.67\pm1.11$ & 1 & \object{\ignore{HIP }107516} & EP Aqr & L & $7.39\pm1.19$ & 5 & $5.98\pm0.90$ & 0 \\
\object{\ignore{HIP } 81835} & S Dra & SRB & $2.39\pm0.75$ & 5 & $2.46\pm0.69$ & 1 & \object{\ignore{HIP }108183} & V413 Cyg & LB & $2.25\pm1.08$ & 5 & $1.02\pm0.84$ & 30 \\
\object{\ignore{HIP } 82912} & RR Sco & M & $2.84\pm1.30$ & 5 & $3.10\pm1.16$ & 4 & \object{\ignore{HIP }108588} & OU Peg & LB & $1.79\pm0.92$ & 5 & $2.40\pm0.88$ & 4 \\
\object{\ignore{HIP } 84213} & TT Dra & SRB & $1.05\pm0.90$ & 5 & $0.82\pm0.84$ & 43 & \object{\ignore{HIP }109070} & SV Peg & SRB & $5.10\pm1.10$ & 5 & $2.53\pm1.03$ & 9 \\
\object{\ignore{HIP } 84346} & V438 Oph & SRB & $-1.25\pm1.41$ & 5 & $-0.04\pm1.20$ & 98 & \object{\ignore{HIP }109089} & RZ Peg & M & $3.54\pm1.36$ & V & $3.23\pm1.23$ & 9 \\
\object{\ignore{HIP } 84833} & V656 Her & LB: & $6.90\pm0.79$ & 5 & $6.55\pm0.71$ & 0 & \object{\ignore{HIP }110146} & X Aqr & M & $-4.01\pm5.73$ & 5 & $1.88\pm3.77$ & 76 \\
\object{\ignore{HIP } 85617} & TW Oph & SRB & $3.57\pm1.34$ & 5 & $1.12\pm0.97$ & 52 & \object{\ignore{HIP }110478} & $\pi^1$ Gru & L & $6.54\pm1.01$ & 5 & $5.01\pm0.95$ & 0 \\
\object{\ignore{HIP } 86836} & Z Oct & M & $-1.18\pm3.04$ & 5 & $1.02\pm1.95$ & 68 & \object{\ignore{HIP }110948} & V410 Lac & LB & $1.28\pm1.06$ & 5 & $2.16\pm0.89$ & 6 \\
\object{\ignore{HIP } 87063} & SX Sco & L & $2.75\pm1.42$ & 5 & $2.47\pm1.13$ & 11 & \object{\ignore{HIP }111043} & $\delta^2$ Gru & LB: & $10.04\pm1.11$ & V & $10.08\pm0.96$ & 0 \\
\object{\ignore{HIP } 87190} & V337 Her & SRB & $3.12\pm0.76$ & 5 & $2.34\pm0.75$ & 2 & \object{\ignore{HIP }112784} & SX Peg & M & $2.12\pm2.99$ & 5 & $4.81\pm2.28$ & 11 \\
\object{\ignore{HIP } 88397} & OU Her & LB & $-0.58\pm1.92$ & 5 & $0.12\pm1.62$ & 95 & \object{\ignore{HIP }112868} & AF Peg  & SRB & $0.93\pm1.55$ & 5 & $1.93\pm1.35$ & 26 \\
\object{\ignore{HIP } 88838} & VX Sgr & SRC & $3.03\pm1.95$ & 5 & $1.86\pm1.75$ & 53 & \object{\ignore{HIP }112961} & $\lambda$ Aqr  & L & $8.33\pm1.13$ & V & $8.82\pm0.91$ & 0 \\
\object{\ignore{HIP } 89568} & RY Oph & M & $3.17\pm1.95$ & 5 & $3.46\pm1.54$ & 16 & \object{\ignore{HIP }113131} & HR Peg & SRB & $3.37\pm0.94$ & 5 & $2.05\pm0.79$ & 5 \\
\object{\ignore{HIP } 89739} & RS Tel & RCB & $3.00\pm2.98$ & 5 & $4.74\pm2.78$ & 21 & \object{\ignore{HIP }114017} & ER Aqr & LB & $4.10\pm1.26$ & 5 & $4.40\pm1.12$ & 0 \\
\object{\ignore{HIP } 89980} & V4028 Sgr & SR: & $2.80\pm1.11$ & 5 & $2.76\pm1.03$ & 4 & \object{\ignore{HIP }114757} & TY And & SRB & $0.87\pm1.32$ & 5 & $1.37\pm1.20$ & 43 \\
\object{\ignore{HIP } 90709} & SS Sgr & SRB & $5.92\pm2.86$ & V & $6.21\pm2.77$ & 5 & \object{\ignore{HIP }115188} & W Peg & M & $3.46\pm1.38$ & 5 & $3.36\pm1.14$ & 4 \\
\object{\ignore{HIP } 90883} & T Lyr & L & $1.58\pm0.75$ & 5 & $2.13\pm0.64$ & 1 & \object{\ignore{HIP }116018} & & LC & $1.64\pm1.05$ & 5 & $1.25\pm0.92$ & 30 \\
\object{\ignore{HIP } 93605} & SU Sgr & SR & $3.14\pm1.44$ & V & $1.45\pm1.36$ & 25 & \object{\ignore{HIP }116681} & ST And & SRA & $5.50\pm2.83$ & 5 & $-0.07\pm1.79$ & 97 \\
\object{\ignore{HIP } 93820} & R Aql & M & $4.73\pm1.19$ & 5 & $3.96\pm0.94$ & 1 & \object{\ignore{HIP }118188} & R Cas & M & $9.37\pm1.10$ & 5 & $10.04\pm1.10$ & 0 \\
\hline
\end{tabular}
\end{table*}

\begin{table}[htb]
\caption{\label{Tab:ReddestVIM}The 19 stars flagged as VIM in the
  Hipparcos catalogue, and  with $(V-I)_{H75} > 5.0$.
The $(V-I)_{H75}$ index is compared to the best value of $V-I$
available at the time of the publication of the Hipparcos and Tycho
Catalogues (field H40) and to the range of $V-I$ used in the
chromaticity corrections. Among these 19 stars, only HIP~2215,
HIP~68815 and HIP~93605 remain VIM at the 0.27\% confidence level
after our improved reprocessing}
\setlength{\tabcolsep}{1.3mm}
\begin{tabular}{lccclccc}\hline
HIP & H75 & H40 & $V-I$ & HIP & H75 & H40 & $V-I$ \\ \hline 2215 &
5.77 & 3.91 & 3.68--3.80 & 87190 & 5.34 & 3.74 & 3.35--3.74 \\ 12193 &
5.92 & 1.15 & 2.26--4.81 & 93605 & 6.66 & 4.32 & 4.25--4.42 \\ 44995 &
5.15 & 5.15 & 3.95--4.71 & 97629 & 6.13 & 6.13 & 4.04--6.28 \\ 47886 &
5.15 & 3.41 & 4.49--5.80 & 98031 & 5.34 & 4.94 & 4.33--5.05 \\ 48036 &
9.03 & 9.03 & 3.97--5.92 & 99082 & 5.39 & 4.87 & 4.51--4.79 \\ 67419 &
9.29 & 5.36 & 4.86--5.56 & 100605 & 5.34& 4.49 & 4.25--4.48 \\ 68815 &
5.88 & 4.10 & 3.64--4.02 & 108588 & 5.51& 5.51 & 3.64--3.79 \\ 70969 &
7.60 & 4.60 & 4.36--4.65 & 110478 & 6.51& 4.65 & 3.99--4.30\\ 71802 &
5.79 & 4.09 & 3.76--4.03 & 118188 & 5.34& 5.34 & 4.02--6.03 \\ 77284 &
5.34 & 5.34 & 3.77--4.11 & \\ \hline
\end{tabular}
\end{table}

%
\section{Discussion}\label{Sect:Results}
%

There are several possible ways to check the validity of the new
results:
\begin{enumerate}
\item The revised parallax may be compared to that estimated by
  \citet{Mennessier-2001:a} using a maximum likelihood approach
  incorporating kinematic and photometric data, distribution functions 
  and a model of the Galaxy.
\item The revised parallax may be compared to estimates based on the
  period-luminosity relationship for Mira variables.
\item A direct confirmation that VIM solutions indeed correspond to
  visual binaries should be seeked. Conversely, it should be checked
  whether the known visual binaries among LPVs indeed  behave as VIM when
  they are not resolved bt Hipparcos.
\end{enumerate}
Each of these items is discussed in turn in the remainder of this
section.

\subsection{Comparison with Mennessier \& Luri parallaxes}
\label{Sect:LPV}

\begin{figure}[htb]
\resizebox{\hsize}{!}{\includegraphics{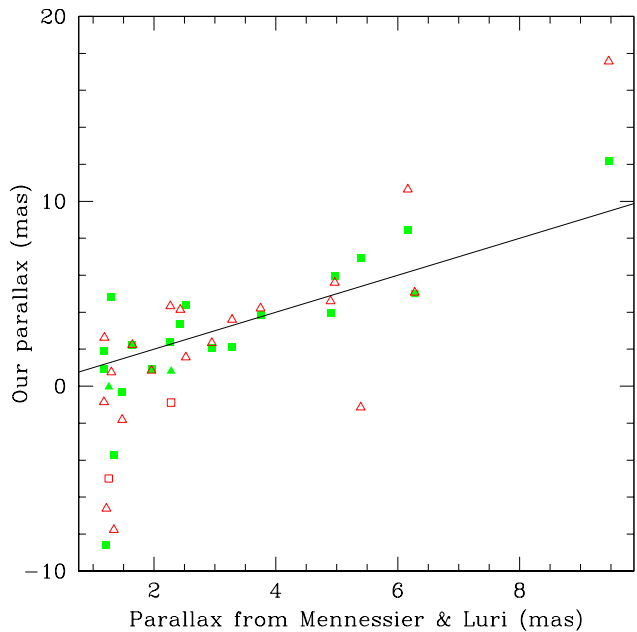}}
\caption[]{\label{Fig:Luri}Comparison of the parallaxes from 
\cite{Mennessier-2001:a} with those obtained with a VIM model
(triangles) or with a single-star model (squares).  
Filled symbols represent the adopted model at the 0.27\% confidence
level (i.e., for a VIM solution, $F_D>3.44$ in
Eq.~(\protect\ref{Eq:FD})). The diagonal is represented by a solid line.
}
\end{figure}

\citet{Mennessier-2001:b} used a maximum likelihood
method to estimate distances of long-period variable stars, based on
kinematic and photometric data, and assuming {\it a priori} functional 
relationships describing the Galaxy and the stellar sample (exponential 
distribution of the number of stars perpendicular to the galactic
plane, gaussian distribution of the absolute magnitudes, Schwarzschild 
velocity ellipsoid). 
Among the stars studied by \citet{Mennessier-2001:a}, 23 also appear
in our sample of confirmed Hipparcos VIM stars for which the epoch $V-I$ indices are now available.
The absolute $K$ magnitudes listed by \citet{Mennessier-2001:a} have
been converted
to parallaxes using apparent $K$ magnitudes from the literature
\citep[see ][]{CaInOb,Whitelock-2000:a}, 
neglecting interstellar extinction and circumstellar reddening. These
`astrophysical' 
parallaxes are compared in Fig.~\ref{Fig:Luri} with the reprocessed single-star
or VIM solutions. Except for parallaxes lower than 2~mas, where the
Hipparcos data are of too poor a quality to derive meaningful
parallaxes, the single-star solution is generally closer to
the estimated `astrophysical' parallax than the VIM solution. Rejecting the VIM solution in those cases requires a much 
more severe confidence level than the 10\% level adopted so far. It is 
found that a level at 0.27\% (corresponding to $F_D > 3.44$) prevents
from adopting the VIM solutions lying far off the diagonal on
Fig.~\ref{Fig:Luri}. Incidentally, we note that this level is the same 
as that used in the Hipparcos catalogue to accept a DMSA/G solution
(i.e., a model with an acceleration term). At the 0.27\% confidence
level, only three VIM solutions pass the test: HIP~36669, 79233, and 94706. 
To our knowledge, none of these is known to be a visual binary, 
but all three have small ($< 1$~mas), non-meaningful parallaxes.

With this much higher confidence level, only 27
VIM solutions are retained (as listed in Table~\ref{Tab:VIM}), 
i.e., just 14\% of the original sample (Fig.~\ref{Fig:VIM2}c,d).
We will show in Sect.~\ref{Sect:binary} that this more stringent
confidence level appears fully justified also for pre-main sequence binaries. 

To conclude, it should be noted that the  \citet{Mennessier-2001:a}
analysis provides a way to evaluate the accuracy of our 
revised parallax  for HIP~99082, which differs by more than $3\sigma$
from the Hipparcos value. 
Based on their maximum-likelihood estimate for $M_{\rm K}$ ($-8.03$)
and the 2~$\mu$m sky-survey $K$
magnitude \citep[$-1.41$;][]{IRC}, a parallax 
$\varpi = 4.7$~mas is derived, 
in good agreement with our reprocessed value of $\varpi = 5.02 \pm
0.97$~mas, which both differ significantly from 
the Hipparcos value 
$\varpi_{\rm HIP} = 1.81\pm0.84$~mas.

%
%

\subsection{Period-Luminosity relation}

The period-luminosity (P-L) relation proposed by \citet{Whitelock-2000:b}
for long-period variables may be used to evaluate the quality of our 
revised parallaxes, although this argument to some extent is circular, since 
the Whitelock \& Feast's P-L relation is based on Hipparcos data.

Nevertheless, the P-L relation provides a very useful diagnostic for 
HIP~65835 (= R~Hya). The Hipparcos VIM solution yields $\varpi = 
1.62 \pm 2.43$~mas, to be compared with $8.44\pm1.00$ for our
reprocessed {\it single-star} model, which lies many $\sigma$ away
from the Hipparcos value! Which value then is the correct one?
Based on their P-L relation and new $K$
magnitudes, \citet{Whitelock-2000:a} derive a distance of 
0.14~kpc ($\varpi=7.14$ mas) for R~Hya which agrees (to within
$1.3\sigma$) with our new parallax.   

Similar cases are HIP~1901 (= R~And) and HIP~77027 (= BG~Ser), 
with $\varpi_{\rm HIP,VIM} = -0.06\pm 6.49$~mas and $-3.78\pm2.74$~mas,
as compared to $6.96\pm3.63$~mas and $2.67\pm2.05$~mas, respectively, 
for our revised single-star 
solutions. For R~And, the P-L relation yields $\varpi = 2.3$~mas
\citep{Whitelock-2000:a}, $1.3\sigma$ away from our revised estimate,
and for BG Ser, 1.92~mas, or $0.4 \sigma$ away. 
HIP~1834 (= T~Cas) is another case where the revised parallax
($\varpi = 3.08\pm0.90$) is in agreement with both the P-L distance
estimate ($\varpi = 3.45$~mas) and the \citet{Mennessier-2001:a} value 
\citep[$\varpi = 2.48$~mas, from $M_{\rm K}$ and the 2~$\mu$m sky-survey $K$
magnitude, ][]{IRC}, despite being very different from $\varpi_{\rm
  HIP} = 0.59\pm 1.07$~mas. 

Thus when large discrepancies exist between the reprocessed and the
Hipparcos parallaxes, the reprocessed parallax appears quite satisfactory.
A more thorough re-investigation of the P-L relation will be presented 
in a separate paper \citep{Knapp-2002:b}.

\subsection{Known visual binaries}
\label{Sect:binary}

\begin{figure}[htb]
\resizebox{\hsize}{!}{\includegraphics{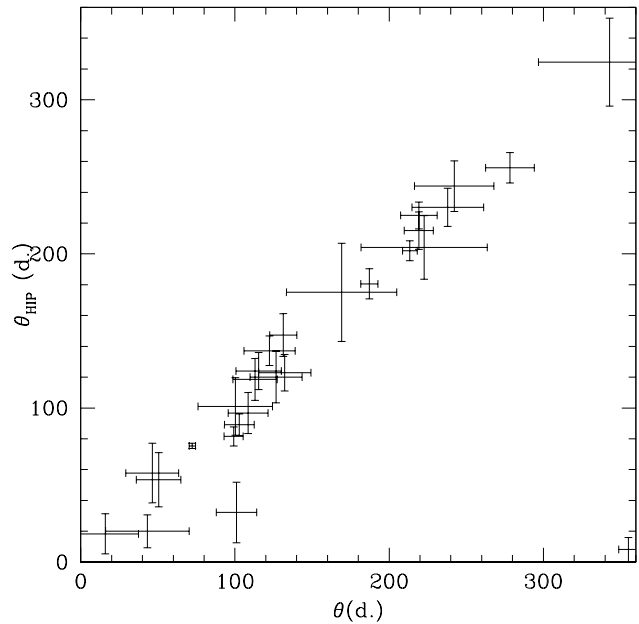}}
\caption[]{\label{Fig:theta}Position angles with the old and new VIM
  solutions.  HIP~67410 is the only case where the two angles are different
but the parallax is small anyway.}
\end{figure}

The VIM model does not provide the angular separation between the
primary and secondary, unless the magnitude of the latter is
known. Therefore
only the position angle ($\theta$) can be derived in all cases as
\begin{equation}
\tan \theta={D_{\alpha^*}\over D_{\delta}},
\end{equation}
where $D_{\alpha^*}$ and $D_{\delta}$ have been defined in relation
with Eq.~(\ref{Eq:xi}). 
Our rederived value for the position angle and the Hipparcos one agree
fairly well (Fig.~\ref{Fig:theta}), except for HIP~67410.

\subsubsection{Pre-main sequence stars}

\citet{Bertout-1999:a} critically evaluated the eight pre-main sequence
stars flagged as VIM in the Hipparcos catalogue, with 
$F_D$ ranging from 2.23 to 9.08.  
Only four of these were known to be binaries before Hipparcos. 
Among them, only \object{HIP 20777} = DF Tau ($F_D=2.79$) would be
discarded with the threshold on $F_D$ set to 3.44.
\citet{Bertout-1999:a} indeed confirm that the VIM solution is not
entirely satisfactory.  On the other hand, at that confidence level, 
only \object{HIP 100289} =  V1685 Cyg
($F_D=3.96$) would remain a newly discovered binary.  In conclusion,
the more stringent criterion on $F_D$ suggested from the study of
long-period variables (Sect.~\ref{Sect:LPV}) 
is perfectly acceptable for
pre-main sequence stars as well, 
since most of the genuine binaries are accepted VIM solutions.

\subsubsection{Long-period variables}\label{Sect:LPVspecific}

Among the 188 objects listed in Table~\ref{Tab:VIM}, 
not many are known to be genuine close visual binaries \citep[see][ for 
a complete list of binaries involving long-period
variables]{Jorissen-2003:a}. 

There first is a group a five close visual binaries which are indeed
accepted as VIM:
\begin{description}
\item{\object{HIP 91389} (= X~Oph = WDS 18384+0850)}. Although X~Oph does not appear in
Table~\ref{Tab:VIM} (because  the Hipparcos catalogue flags 
it as a stochastic solution DMSA/X), it is  
almost a textbook case. It has a variation range $Hp = 5.8$ to 8.1
(5th to 95th percentile), and a $V = 8.6$ companion located $0\farcs4$ 
away at a position angle of $131^\circ$. The confidence level of our
VIM solution is very high ($F_D = 72.4$), and is in very good
agreement with the ground based data, since it yields a position angle 
of $147^\circ$ and a separation of $0\farcs328$. The only worry about
this solution is its negative parallax, and the large residuals
(20~mas), indicative that there is some other effect present, possibly 
the orbital motion \citep[although the estimated orbital period -- 485~y -- is quite long with 
respect to the duration of the Hipparcos mission;][]{Hartkopf-2001:a}. 

\item{HIP~46806 (= R~Car = WDS 09322-6247)}. A close visual binary with a
$V=11.3$ companion located at $\rho =
1\farcs8$ and $\theta = 140^\circ$, with an accepted VIM
reprocessed solution ($F_D = 6.48$). Despite the high confidence level 
of the VIM solution, it yields a position 
angle $\theta = 355^\circ\pm7^\circ$ in disagreement with the
ground-based solution.  

\item{HIP~79233 (=RU~Her)}. This star  
 is suspected to be a binary by \citet{Herbig-1965:a} who notes that
 {\it the 
absorption features are remarkably weak, as though veiled, the
spectrum may be composite}.  With $F_D=3.53$, the VIM solution is indeed
adopted.

\item{HIP~93605 (= SU~Sgr = WDS 19037-2243)}. This star is listed by
\citet{Proust-1981:a} as having a $V=13.3$ companion at $\rho =
1\farcs1$ and $\theta = 239^\circ$. The semiregular variable SU~Sgr
varied between $Hp = 7.3$ and 7.9 (5th and 95th percentiles, fields
H49 and H50) during the Hipparcos mission. Using the formulae of
\cite{Wielen-1996}, we find that, in such conditions, the
photocenter should exhibit a back and forth motion with a total
amplitude of 3.3~mas. This motion is probably marginally detectable by 
Hipparcos, except that the companion may be below the photometric
detection threshold. The position angle of the companion with respect
to the variable star obtained from the VIM solution ($\theta =
102^\circ \pm 7^\circ$) is, however, not consistent with the
ground-based position angle ($\theta =
239^\circ$), thus casting a doubt on the
reality of the VIM solution.  

\item{HIP~94706 (=T~Sgr)}. This star is also noted by
\citet{Herbig-1965:a} as having a composite spectrum, later confirmed
by \citet{Culver-1975:a}. Although it is flagged as VIM in the
Hipparcos catalogue, both the Hipparcos and our reprocessed parallaxes 
(also obtained within the framework of a VIM model) are negative by a
large amount. There must thus be another effect disturbing the VIM
solution, perhaps an orbital motion (the orbital period is still unknown).
\end{description}
The discrepancy between ground-based and Hipparcos results for HIP~46806 and 93605 may result from the 1.2\arcsec periodicity of the focal-plane grid, thus limiting the effectiveness of the VIM model to systems with separations below 0.5\arcsec.

There is then a group of 3 stars which are close visual binaries, but not
accepted as VIM.
The two carbon semiregular variable stars with composite spectra 
HIP~27135 (= TU~Tau) and
HIP~86873 \citep[=SZ~Sgr;][]{Olson-1975} 
were neither flagged as VIM by the Hipparcos catalogue, nor by our 
reprocessing.   
The VIM nature of HIP~110478 (= $\pi^1$~Gru) is not confirmed
by our reprocessing, but the companion is indeed rather far away ($2.7''$).
We may thus conclude that the number of genuine binaries not detected
as VIM is indeed small.

Finally, several stars not accepted as VIM in  Table~\ref{Tab:VIM} 
are seen as single, as it should be, 
by speckle observations having a resolution of about 50~mas: 
HIP~63950 \citep{Mason-1999:b}, HIP~1834, 35045, 95173, 109070,
110948, 112868, and 115188 \citep{Mason-2001:a}.  

HIP~36288 (= Y~Lyn) is the only puzzling case, since despite
being confirmed as VIM by our reprocessing, it
could not be resolved by \citet{Mason-2001:a}. The only noteworthy
property of this star possibly relevant in the present context is its
spatial extension in the IRAS 60 and 100~$\mu$m passbands \citep{Young-1993:a}.

%
%

%
\section{Conclusion}\label{Sect:Conclusion}
%

Instantaneous $V-I$ indices derived from a color transformation based on 
$Hp - V_{T2}$ have been used to improve the chromaticity correction of 
red and variable stars. All stars flagged as Variability-Induced Mover 
(VIM) in the Hipparcos
catalogue have been reprocessed.
It turns out that about 85\% of those objects can now be
satisfactory modeled with the basic 5-parameter, single-star solution.
This result explains why many of the Hipparcos VIM solutions did not
receive confirmation of their binary nature from speckle observations.

\begin{acknowledgements}
We thank L.~Lindegren for valuable suggestions and for providing us
the unpublished NDAC chromaticity constants, C.~Dettbarn for technical details
about the actual processing of the VIM by the FAST consortium,
and C.~Fabricius for his help with the Tycho-2 photometry.  This research
was supported in part by ESA/PRODEX C15152/01/NL/SFe(IC) and by NASA
via grant NAG5-11094.
\end{acknowledgements}

\bibliographystyle{aa} 
\bibliography{articles,books}

\begin{thebibliography}{23}
\expandafter\ifx\csname natexlab\endcsname\relax\def\natexlab#1{#1}\fi

\bibitem[{{Bertout} {et~al.}(1999){Bertout}, {Robichon}, \&
  {Arenou}}]{Bertout-1999:a}
{Bertout}, C., {Robichon}, N., \& {Arenou}, F. 1999, A\&A, 352, 574

\bibitem[{{Culver} \& {Ianna}(1975)}]{Culver-1975:a}
{Culver}, R.~B. \& {Ianna}, P.~A. 1975, ApJ, 195, L37

\bibitem[{{ESA}(1997)}]{Hipparcos}
{ESA}. 1997, The Hipparcos and Tycho Catalogues (ESA SP-1200)

\bibitem[{{Gezari} {et~al.}(1999){Gezari}, {Pitts}, \& {Schmitz}}]{CaInOb}
{Gezari}, D.~Y., {Pitts}, P.~S., \& {Schmitz}, M. 1999, Catalog of Infrared
  Observations, 5th edn.

\bibitem[{{Hartkopf} {et~al.}(2001){Hartkopf}, {Mason}, \&
  {Worley}}]{Hartkopf-2001:a}
{Hartkopf}, W.~I., {Mason}, B.~D., \& {Worley}, C.~E. 2001, AJ, 122, 3472

\bibitem[{{Herbig}(1965)}]{Herbig-1965:a}
{Herbig}, G.~H. 1965, Kleine Ver\"off. Remeis-Sternwarte, 164

\bibitem[{{H\o g} {et~al.}(2000){H\o g}, {Fabricius}, {Makarov}, {Urban},
  {Corbin}, {Wycoff}, {Bastian}, {Schwekendiek}, \& {Wicenec}}]{Hog-2000:a}
{H\o g}, E., {Fabricius}, C., {Makarov}, V.~V., {et~al.} 2000, A\&A, 355, L27

\bibitem[{{Jorissen}(2003)}]{Jorissen-2003:a}
{Jorissen}, A. 2003, in Asymptotic giant branch stars, ed. H.~{Habing} \&
  H.~{Olofsson} (New York: Springer Verlag)

\bibitem[{{Kholopov} {et~al.}(1998){Kholopov}, {Samus}, {Frolov}, {Goranskij},
  {Gorynya}, {Karitskaya}, {Kazarovets}, {Kireeva}, {Kukarkina}, {Kurochkin},
  {Medvedeva}, {Pastukhova}, {Perova}, {Rastorguev}, \&
  {Shugarov}}]{Kholopov-1998:a}
{Kholopov}, P.~N., {Samus}, N.~N., {Frolov}, M.~S., {et~al.} 1998, Combined
  General Catalogue of Variable Stars, 4th edn.

\bibitem[{{Knapp} {et~al.}(2002){Knapp}, {Pourbaix}, {Platais}, \&
  {Jorissen}}]{Knapp-2002:b}
{Knapp}, G.~R., {Pourbaix}, D., {Platais}, I., \& {Jorissen}, A. 2002, A\&A,
  (in preparation)

\bibitem[{{Lindegren} {et~al.}(1997){Lindegren}, {Mignard}, {S\"oderhjelm},
  {Badiali}, {Bernstein}, {Lampens}, {Pannunzio}, {Arenou}, {Bernacca},
  {Falin}, {Froeschl\'e}, {Kovalevsky}, {Martin}, {Perryman}, \&
  {Wielen}}]{Lindegren-1997:a}
{Lindegren}, L., {Mignard}, F., {S\"oderhjelm}, S., {et~al.} 1997, A\&A, 323,
  L53

\bibitem[{{Mason} {et~al.}(2001){Mason}, {Hartkopf}, {Holdenried}, \&
  {Rafferty}}]{Mason-2001:a}
{Mason}, B.~D., {Hartkopf}, W.~I., {Holdenried}, E.~R., \& {Rafferty}, T.~J.
  2001, AJ, 121, 3224

\bibitem[{{Mason} {et~al.}(1999){Mason}, {Martin}, {Hartkopf}, J., {Germain},
  {Douglas}, {Worley}, {Wycoff}, {Brumelaar}, \& {Franz}}]{Mason-1999:b}
{Mason}, B.~D., {Martin}, C., {Hartkopf}, W.~I., {et~al.} 1999, AJ, 117, 1890

\bibitem[{{Mennessier} \& {Luri}(2001)}]{Mennessier-2001:a}
{Mennessier}, M.~O. \& {Luri}, X. 2001, A\&A, 380, 198

\bibitem[{{Mennessier} {et~al.}(2001){Mennessier}, {Mowlavi}, {Alvarez}, \&
  {Luri}}]{Mennessier-2001:b}
{Mennessier}, M.~O., {Mowlavi}, N., {Alvarez}, R., \& {Luri}, X. 2001, A\&A,
  374, 968

\bibitem[{{Neugebauer} \& {Leighton}(1969)}]{IRC}
{Neugebauer}, G. \& {Leighton}, R.~B. 1969, Two-Micro Sky Survey Catalogue
  (NASA SP-3047)

\bibitem[{{Olson} \& {Richer}(1975)}]{Olson-1975}
{Olson}, B.~I. \& {Richer}, H.~B. 1975, ApJ, 200, 88

\bibitem[{{Platais} {et~al.}(2002){Platais}, {Pourbaix}, {Jorissen}, {Makarov},
  {Berdnikov}, {Samus}, {Lloyd {E}vans}, {Lebzelter}, \&
  {Sperauskas}}]{Platais-2002:a}
{Platais}, I., {Pourbaix}, D., {Jorissen}, A., {et~al.} 2002, A\&A, (accepted)

\bibitem[{{Proust} {et~al.}(1981){Proust}, {Ochsenbein}, \&
  {Pettersen}}]{Proust-1981:a}
{Proust}, D., {Ochsenbein}, F., \& {Pettersen}, B.~R. 1981, A\&AS, 44, 179

\bibitem[{{Whitelock} \& {Feast}(2000)}]{Whitelock-2000:b}
{Whitelock}, P. \& {Feast}, M. 2000, MNRAS, 319, 759

\bibitem[{{Whitelock} {et~al.}(2000){Whitelock}, {Marang}, \&
  {Feast}}]{Whitelock-2000:a}
{Whitelock}, P., {Marang}, F., \& {Feast}, M. 2000, MNRAS, 319, 728

\bibitem[{{Wielen}(1996)}]{Wielen-1996}
{Wielen}, R. 1996, A\&A, 314, 679

\bibitem[{{Young} {et~al.}(1993){Young}, {Phillips}, \& {Knapp}}]{Young-1993:a}
{Young}, K., {Phillips}, T.~G., \& {Knapp}, G.~R. 1993, ApJS, 86, 517

\end{thebibliography}
\end{document}